
\tolerance=10000
\magnification=1200
\raggedbottom

\baselineskip=15pt
\parskip=1\jot

\def\sk{\vskip 3\jot}

\def\heading#1{\vskip3\jot{\noindent\bf #1}}
\def\label#1{{\noindent\it #1}}


\def\ref#1;#2;#3;#4;#5.{\item{[#1]} #2,#3,{\it #4},#5.}
\def\refinbook#1;#2;#3;#4;#5;#6.{\item{[#1]} #2, #3, #4, {\it #5},#6.} 
\def\refbook#1;#2;#3;#4.{\item{[#1]} #2,{\it #3},#4.}


\def\({\bigl(}
\def\){\bigr)}


\def\et{\eta}
\def\th{\vartheta}

\def\ka{\kappa}

\def\rh{\varrho}
\def\si{\sigma}

\def\me{\omega}



\def\Ex{{\rm Ex}}

\def\fp#1{\{#1\}}

\def\elo#1{\Ex[ \#{#1} ]}

{
\baselineskip=12.35pt
\pageno=0
\nopagenumbers
\rightline{\tt zpx.tex}
\vskip1in

\centerline{\bf Asymptotic Analysis of Run-Length Encoding}
\vskip0.5in

\centerline{Nabil Zaman}
\centerline{\tt nzaman@g.hmc.edu}
\sk

\centerline{Nicholas Pippenger}
\centerline{\tt njp@math.hmc.edu}
\sk

\centerline{Harvey Mudd College}
\centerline{301 Platt Bouldevard}
\centerline{Claremont, CA 91711}
\vskip0.5in

\noindent{\bf Abstract:}
Gallager and Van Voorhis have found optimal prefix-free codes $\ka(K)$ 
for a random variable $K$ that is geometrically distributed: $\Pr[K=k] = p(1-p)^k$ for $k\ge 0$.
We determine the asymptotic behavior of the expected length $\elo{\ka(K)}$ of these codes as 
$p\to 0$:
$$\elo{\ka(K)} = \log_2 {1\over p} + \log_2 \log 2 + 2 
+ f\left(\log_2 {1\over p} + \log_2 \log 2\right) + O(p),$$
where
$$f(z) = 4\cdot 2^{-2^{1-\fp{z}}} - \fp{z} - 1,$$
and $\fp{z} = z - \lfloor z\rfloor$ is the fractional part of $z$.
The function $f(z)$ is a periodic function (with period $1$) that exhibits small oscillations (with magnitude less than
$0.005$) about an even smaller average value (less than $0.0005$).
\vfill\eject
}

\heading{1. Introduction}

In 1975, Gallager and Van Voorhis [G1] (building on prior work by Golomb [G2]) found optimal prefix-free codes for a geometrically distributed random variable; that is, a random variable $K$ such that
$$\Pr[K=k] = p(1-p)^{k} \eqno(1.1)$$
for $k\ge 0$, where $0<p<1$ is a parameter.
(This problem is sometimes referred to as the ``run-length encoding'' problem, since the number of $0$s between consecutive $1$s in a sequence of independent and identically distributed Bernoulli random variables is geometrically distributed.)
Their result shows that the optimal codes $\ka(K)$ have expected codeword length 
$\elo{\ka(K)}]$ close, but not equal, to the lower bound given by the entropy
$$\eqalignno{
\elo{\ka(K)}
&\ge H(K) \cr
&= - \sum_{k\ge 0} p(1-p)^{k} \log_2 \(p(1-p)^{k}\) \cr
&=   \log_2 {1\over p} + {1-p\over p}\log_2 (1-p) \cr
&=  \log_2 {1\over p} + \log_2 e + O(p), &(1.2)\cr
}$$ 
where $\log_2 e = 1.442\ldots\,$.

In Section 2,
we shall show that 
$$\elo{\ka(K)} = \log_2 {1\over p} + \log_2 \log 2 + 2 
+ f\left(\log_2 {1\over p} + \log_2 \log 2\right) + O(p), \eqno(1.3)$$
where the function $f(z)$ is a bounded
periodic function of $z$ with period $1$.
Specifically, 
$$f(z) = 4\cdot 2^{-2^{1-\fp{z}}} - \fp{z} - 1,$$
with
$\{z\} = z - \lfloor z\rfloor$ denoting the fractional part of $z$.
The function $f(z)$ exhibits small oscillations about its average 
value $\me = \int_0^1 f(z)\,dz = 4(\log_2 e)\(E_1(\log 2) - E_1(2\log 2)\) - 3/2 = 0.0004547\ldots\,$,
where $E_1(y) = \int_y^\infty (e^{-x}  /  x) \, dx$.
It assumes its largest value of $f(z_1) = 0.004195\ldots\,$ at 
$z_1 = 1 + \log_2\log 2 - \log_2 x_1 = 0.7680\ldots\,$, where $x_1 = 0.8140\ldots$ is the smaller solution of the equation $xe^{-x} = (\log_2 e)/4$, 
and its smallest value of $f(z_0) = -0.003438\ldots\,$ at 
$z_0 = 1 + \log_2 \log 2 - \log_2 x_0 = 0.1934\ldots\,$, where $x_0 = 1.2123\ldots$ is the larger solution of that equation.
Comparing this result with (1.2), we see that the average redundancy of the optimal code is
$\log_2 \log 2 + 2 + \me - \log_2 e = 0.02899\ldots\,$.
\sk

\heading{2. Run-Length Encoding}

According to Gallager and Van Voorhis [G1], the optimal prefix-free binary codes for a geometric random variable $K$, distributed according to (1.1), can be constructed as follows.
Set
$$m = \left\lceil {\log (2-p)\over -\log(1-p)} \right\rceil. \eqno(2.1)$$
Divide $K$ by $m$ to obtain a quotient $S\ge 0$ and a remainder $0\le R\le m-1$:
$$K = Sm + R.$$
The distribution of $S$ is geometric with parameter $q = 1-(1-p)^m$; that is
$$\Pr[S=s] = (1-q)q^s$$
for $s\ge 0$.
The distribution of $R$ is ``truncated geometric'':
$$\Pr[R=r] = {(1-p)^r \over 1-(1-p)^m} \eqno(2.2)$$
for $0\le r\le m-1$.
We shall take optimal prefix-free codes $\si(S)$ for $S$ and $\rh(R)$ for $R$,
and concatenate them (as strings) to obtain an optimal code $\ka(K)=\si(S)\,\rh(R)$.

Since (2.1) implies $q\ge (3-\sqrt{5})/2$, an optimum prefix-free code for the quotient $S$ is 
$\si(S) = 1^S\,0$, and the expected length of this code is
$$\elo{\si(S)} = \Ex[S] + 1 = {1\over q} = {1\over 1-(1-p)^m}. \eqno(2.3)$$

The optimum prefix-free code for the remainder $R$ is a Huffman code $\rh(R)$ (see Huffman [H]). 
We observe that the ratio between the smallest and the largest of the probabilities given by (2.2) is
$(1-p)^{m-1}$.
Since (2.1) implies that $(1-p)^{m-1} \ge 1/(2-p) > 1/2$, it follows that all of the codewords in this Huffman code must be of at most two consecutive lengths.
(If there were a codeword $\xi$ of length $i$ and two codewords $\et\,0$ and $\et\,1$ each of length $j\ge i+2$, then the probability of $\xi$ would be strictly smaller than the sum of the probabilities of $\et\,0$ and $\et\,1$,
and the code with $\et$ of length $j-1$ and $\xi\,0$ and $\xi\,1$ each of length $i+1$
would have strictly smaller expected length.)
Take 
$$l = \lfloor \log_2 m\rfloor$$
and
$$h = m - 2^l,$$
so that $0\le h\le 2^l - 1$.
Then there will be $2^l - h$ codewords of length $l$ and $2h$ codewords of length $l+1$.
Since the $2h$ longer codewords will have the $2h$ smallest probabilities, the expected codeword length for the Huffman code is
$$\elo{\rh(R)} = l + \sum_{m-2h\le k\le m-1} {(1-p)^k \over 1 - (1-p)^m}
= l + {(1-p)^{m-2h} - (1-p)^m \over 1 - (1-p)^m}.$$
Combining this expression for the expected length of the encoding of $R$ with (2.3) for the expected length of the encoding of $S$, we obtain
$$\eqalignno{
\elo{\ka(K)}
&=  \elo{\si(S)} + \elo{\rh(R)} \cr
&= l + {(1-p)^{m-2h} - (1-p)^m \over 1 - (1-p)^m} + {1\over 1-(1-p)^m} &(2.4) \cr
}$$
for the expected length of the encoding of $K$.

Since $l$ and $h$ are defined in terms of $m$, we shall start by 
eliminating them in favor of 
$$\th = \fp{\log_2 m},$$
the fractional part of $\log_2 m$.
Then $l = \log_2 m - \th$, and from $m = 2^l + h$ we obtain $1 - 2h/m = 2^{1-\th} - 1$.
Substituting these expressions in (2.4) yields
$$\elo{\ka(K)}=  \log_2 m - \th + {(1-p)^{m(2^{1-\th}-1)} - (1-p)^m \over 1 - (1-p)^m} + {1\over 1-(1-p)^m} \eqno(2.5)$$
Since $\th$ is defined in terms of $m$, our next step will be to eliminate $p$ in favor of $m$ by using the relation
$$m = {\log 2\over p} + O(1), \eqno(2.6)$$
which follows from (2.1) and implies
$$p = {\log 2\over m} + O\left({1\over m^2}\right).$$
Thus 
$$(1-p)^m = {1\over 2} + O\left({1\over m}\right).$$
Substituting this expression in (2.5) yields
$$\elo{\ka(K)} = \log_2 m + 2 + f(\log_2 m) + O\left({1\over m}\right), \eqno(2.7)$$
where 
$$f(z) = 4\cdot 2^{-2^{1-\fp{z}}} - \fp{z} - 1.$$
The function $f$ is periodic with period $1$.
It is also continuous (because $\lim_{z\to 1} f(z) = f(0) = 0$), and has a continuous derivative
(because $\lim_{z\to 1} f'(z) = f'(0) = 2(\log 2)^2 - 1$).
These properties, combined with the relation
$$\log_2 m = \log_2 {1\over p} + \log_2 \log 2 + O(p)$$
which follows from (2.6),
allow us to deduce
$$f(\log_2 m) = f\left(\log_2 {1\over p} + \log_2 \log 2\right) + O(p).$$
This in turn allows us to rewrite (2.7)  in terms of $p$ as
$$\elo{\ka(K)} = \log_2{1\over p} + \log_2 \log 2 + 2 + f\left(\log_2{1\over p} + \log_2\log 2\right) + O(p).$$
It remains to examine the properties of the function $f(z)$.
The continuity of the periodic function $f(z)$ ensures that it is bounded, and the continuity of its derivative ensures that its maxima and minima occur at values of $z$ for which the derivative vanishes.
The vanishing of the derivative is given by the equation
$$1 = 4\cdot 2^{-2^{1-\th}} \cdot 2^{1-\th} \, (\log 2)^2.$$
The substitution $x = 2^{1-\th} \, \log 2$ reduces this equation to
$x\,e^{-x} = (\log_2 e) / 4$, from which the numerical results mentioned in Section 1 follow.
The same substitution also reduces the integral $\me = \int_0^1 f(z)\,dz$ for the average value of $f(z)$ to the integral $4(\log_2 e)\int_{\log 2}^{2\log 2} (e^{-x} / x)\, dx$, leading again to the numerical results mentioned in Section 1.

We mention in closing that Wolf [W] has shown how to use some of the optimal prefix-free codes found by Gallager and Van Voorhis to construct asymptotically optimal nested strategies for group testing.
For this problem, the lower bound (applying to all strategies, nested or not) to the expected number of tests per positive individual  has the asymptotic behavior indicated in (1.2).
Thus the gap between (1.3) and (1.2) represents a bound to the possible advantage that non-nested strategies might have over nested ones.
\vfill\eject

\heading{3. References}

\ref G1; R. G. Gallager and D. C.  Van Voorhis;
``Optimal Source Codes for Geometrically Distributed Integer Alphabets'';
IEEE Trans.\ Information Theory; 21:2 (1975) 228--229.

\ref G2; S. W. Golomb;
``Run-Length Encodings'';
IEEE Trans.\ Information Theory; 12:3 (1966) 399--401.

\ref H; D. Huffman;
``A Method for the Construction of Minimum Redundancy Codes'';
Proc.\ IRE; 40 (1952) 1098--1103.

\ref W; J. Wolf;
``Born Again Group Testing: Multiaccess Communications'';
IEEE Trans.\ Information Theory; 31:2 (1985) 185--191.

\bye